\documentclass[twocolumn,twoside]{IEEEtran}

\makeatletter
\let\NAT@parse\undefined
\makeatother

\usepackage{amsmath,amssymb,amsfonts,cuted}
\usepackage[final]{graphicx}
\usepackage{psfrag}
\usepackage{epsfig}
\usepackage[numbers,sort&compress]{natbib}

\usepackage{flushend}
\usepackage{array,booktabs}

\usepackage{amsmath,amssymb,amsfonts,amsthm}
\usepackage[final]{graphicx}
\usepackage{flushend}
\usepackage[numbers,sort&compress]{natbib}

\usepackage{ucs} 
\usepackage[utf8x]{inputenc}
\usepackage[usenames,dvipsnames]{xcolor}
\usepackage{tikz}
\usepackage{tkz-tab}
\usepackage{multirow}
\usepackage{latexsym}
\usepackage{amssymb}
\usepackage{amsmath}
\usepackage{mathrsfs}
\usepackage{amsthm}

\newtheorem{lemma}{Lemma}

\makeatletter
\def\blfootnote{\xdef\@thefnmark{}\@footnotetext}
\makeatother

\begin{document}

\title{Statistical Characterization of Second Order Scattering Fading Channels}
\author{J. Lopez-Fernandez, F.~J. Lopez-Martinez}
\maketitle

\begin{abstract}
We present a new approach to the statistical characterization of the second order scattering fading (SOSF) channel model, which greatly simplifies its analysis. Exploiting the unadvertised fact that the SOSF channel can be seen as a continuous mixture of Rician fading channels, we obtain expressions for its probability density function and cumulative density function that are numerically better-behaved than those available in the literature. Our approach allows for obtaining new results for the SOSF model, such as a closed-form expression for its moment-generating function, as well as the characterization of the average channel capacity. Relevantly, and somehow counterintuitively, we observe that in the presence of a strong line-of-sight (LOS) component, the channel capacity of a LOS plus double-Rayleigh scattered diffuse component is larger than its LOS plus Rayleigh (i.e Rician-like) counterpart.
%
\end{abstract}

\blfootnote{\noindent  J. Lopez-Fernandez and F.J. Lopez-Martinez are with Dpto. Ingenier\'ia de Comunicaciones, Universidad de Malaga, Malaga 29071, Spain. E-mail: \{jlf,fjlopezm\}@ic.uma.es. This work has been funded by the Spanish Government and the European Fund for Regional Development FEDER (projects TEC2014-57901-R and TEC2017-87913-R).
}

\blfootnote{\noindent  This work has been submitted to the IEEE for publication. Copyright may be transferred without notice, after which this version may no longer be accesible).
}

\section{Introduction}

The statistical characterization of the small-scale random fluctuations of the signal amplitude in a wireless communication context has been one of the key problems in the wireless arena \cite{Simon2005}. Because classical models arising from the central limit theorem like Rayleigh or Rice have shown to be insufficient to accurately fit experimental data in many scenarios, a number of more general and sophisticated models have been proposed to better reflect the specific propagation effects that affect the radio signal \cite{Durgin2002,Salo2006,Yacoub2016,Romero2017,Pablo2018}. 

A relevant example of such generalized models is the second-order scattering fading (SOSF) channel \cite{URSI2002,Salo2006}. This model is a particular case of a family of multiple-order scattering fading channels \cite{Salo2006}, which are built from the combination of a finite number of increasing order scattering terms. For the second order case, the received signal is expressed as the combination of a line-of-sight (LOS) component, plus a Rayleigh-diffused scattering component and a double-Rayleigh scattering component. Therefore, the SOSF fading model has a solid motivation from a physical perspective, in the sense that it captures situations (e.g. propagation in the presence of diffracting wedges like rooftops or building corners) on which the signal propagation is affected by the \textit{keyhole} effect \cite{Keyholes2000}. Reported measurements in quite dissimilar wireless environments like indoor \cite{Bandemer2009}, indoor-to-outdoor, outdoor-to-indoor \cite{Bandemer2010}, peer-to-peer \cite{Vinogradov2015}, urban, suburban, forest \cite{URSI2002} or more recently in high speed railway \cite{Zhang2017}, provide experimental support to this claim. Hence, the SOSF fading model is indeed well-suited to recreate a wide variety of propagation effects.

However, similarly to other state-of-the-art fading models in the literature \cite{Durgin2002,Salo2006,Yacoub2016,Romero2017,Pablo2018}, this often comes at the price of an increased mathematical complexity, which ultimately hinders the understanding of how the fading model parameters impact on performance metrics such as outage probability or channel capacity. This drawback is specially accentuated in the case of the SOSF channel, as its probability density function (PDF) and cumulative distribution function (CDF) involve an infinite integration of Bessel functions of the first kind and zero order. The oscillating nature of the integrand poses a significant inconvenient for the numerical evaluation of PDF and CDF \cite{Salo2006,Markham2003}, and further analytical manipulations for performance analysis purposes are also complicated. In turn, to the best of our knowledge the capacity of SOSF channels remains unknown.

In this paper, we propose a different approach to the analysis of SOSF fading channels which completely avoids these drawbacks, and also facilitates the performance analysis of wireless communication systems operating over this otherwise unwieldy fading channel. We show that the SOSF fading channel can be seen as a continuous mixture of Rician fading channels, i.e. it can be expressed in terms of an underlying Rician random variable (RV), conditioned to an exponentially-distributed ancillary RV. Therefore, its PDF and CDF can be expressed in terms of a single integral involving the Rician PDF and CDF, which are considerably better-behaved than the previously available expressions. We also obtain a closed-form expression for the moment generating function (MGF) of the SOSF channel for the first time in the literature. Closed-form expressions for the PDF, CDF and MGF of all the special cases included in the SOSF model are also obtained. Besides, our approach naturally simplifies the performance analysis in this scenario, as readily available results in the literature for the Rician case can be leveraged to analyze the SOSF case by an additional integration over an exponential distribution. This is illustrated by analyzing the average capacity of SOSF channels, for which we observe some insightful effects not previously reported in the literature.

The remainder of the paper is organized as follows: in section \ref{Sec:The system model} we briefly revisit the SOSF physical model. In section \ref{Sec:Statistical analysis} we carry out the statistical characterization of the SOSF model, obtaining a collection of expressions for the PDF, CDF and MGF of the general SOSF fading distribution, and all the special cases derived from it. Section \ref{Application example: Ergodic Capacity} is devoted to analyze the average capacity of the SOSF channel, with a special focus on the high signal-to-noise (SNR) ratio regime. Finally, the main conclusions are drawn in section \ref{Conclusions}.

\emph{Notation:} The expectation and the absolute value of a RV $X$ are denoted as $\mathbb{E}(X)$ and $|X|$ respectively. The notation $X|Y$ will stand for $X$ conditioned to $Y$.  We write $X \sim \mathcal{N}_c(\mu,\sigma^2)$ to denote that $X$ is distributed as a circularly-symmetric complex Gaussian RV with complex mean $\mu$ and variance $\sigma^2$. The symbol ${\buildrel d \over =}$ indicates equality in distribution.

\section{System model}
\label{Sec:The system model}

The received signal under SOSF fading is modeled as a random variable $S$ given by \cite{Salo2006}

\begin {equation}
S=\omega_0 e^{j\phi}+\omega_1G_1+\omega_2 G_2 G_3,
\label{Eq:Modelo_SOSF}
\end{equation}
where $\omega_0 e^{j\phi}$ is the LOS component with weighting factor $\omega_0$, and $\phi$ is a random phase uniformly\footnote{We note that the alternative definition of $\phi$ as a deterministic value as in \cite{Bandemer2009} does not have any impact on the envelope statistics of $S$.} distributed in $[0,2\pi)$. $G_i$ for $i=1,2,3$ are independent random variables distributed as $\mathcal{N}_c(0,1)$ so that the term $\omega_1G_1$ corresponds to a Rayleigh fading component (first-order scattering), whereas the last term $\omega_2 G_2 G_3$ models a double-Rayleigh fading component (second-order scattering). The parameters $\omega_0$, $\omega_1$ and $\omega_2$ are non-negative real-valued constants that determine the relative level of each fading component. Notice that the mean square value of $S$ is given by $\mathbb{E}(|S|^2)=\omega_0^2+\omega_1^2+\omega_2^2$. Without loss of generality, we will consider a normalized SOSF channel with $\omega_0^2+\omega_1^2+\omega_2^2=1$,

The SOSF fading model is often specified by an alternative set of parameters ($\alpha, \beta$), defined as
\begin {equation}
\alpha=\frac{\omega_2^2}{\omega_0^2+\omega_1^2+\omega_2^2},\;\;\;\;\;\beta=\frac{\omega_0^2}{\omega_0^2+\omega_1^2+\omega_2^2}.
\label{Eq:alpha_beta}
\end{equation}
The parameters $(\alpha,\beta)$ are constrained to the triangle $\alpha \geq 0$, $\beta \geq 0$ and $\alpha+\beta \leq 1$.

\begin{figure}[t]
\centering
\includegraphics[width=.9\columnwidth]{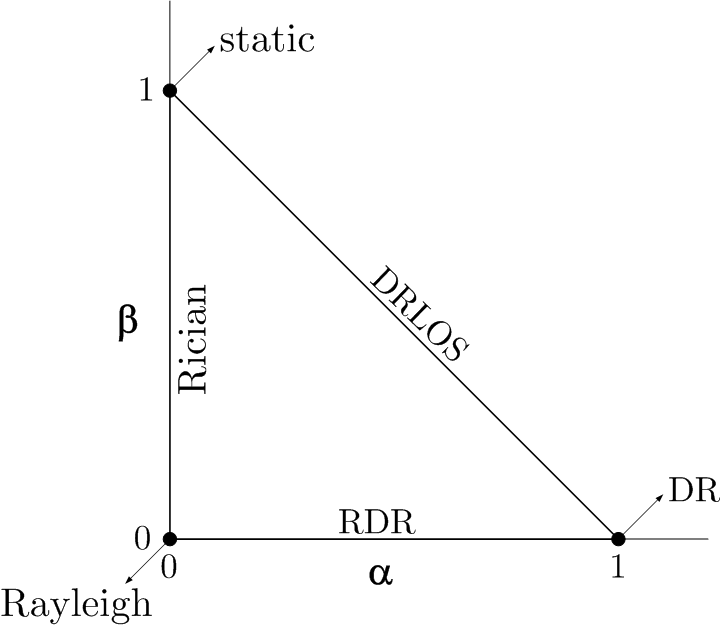}
\centering
\caption{Triangle of permissible values of parameters $\alpha$ and $\beta$ and SOSF distribution designations associated to particular values of the parameters.} 
\label{Fig_Triangle}
\end{figure}

The model described in (\ref{Eq:Modelo_SOSF}) encompasses a number of fading models as special cases, like Rician fading when $\omega_2=0$, with $K$ factor given as $K=\tfrac{\omega_0^2}{\omega_1^2}$, Rayleigh and double-Rayleigh (RDR)\footnote{also denoted as \textit{leaky keyhole} in the literature} fading for $\omega_0=0$, double-Rayleigh and LOS (DRLOS) for $\omega_1=0$, double-Rayleigh (DR) in case $\omega_0=\omega_1=0$, or Rayleigh fading when $\omega_0=\omega_2=0$. Table \ref{Tab:table_fading_types} summarizes all special cases associated to specific values of the set of parameters ($\omega_0,\omega_1,\omega_2$) and ($\alpha, \beta$). Dashes in Table \ref{Tab:table_fading_types} indicate that the corresponding parameter can take any value as long as it meets the constraints of being non-negative and $\omega_0^2+\omega_1^2+\omega_2^2=1$ or equivalently $\alpha+\beta \leq 1$. Fig. \ref{Fig_Triangle} shows the triangular domain of $\alpha$ and $\beta$, on which the three sides and three vertices of the triangle have been labeled with the corresponding SOSF distribution type. 

Let $\gamma$ denote the random instantaneous signal to noise ratio (SNR) of the fading signal and let $\bar \gamma$ denote the average received SNR. 
Thus, we have $\gamma \propto |S|^2$, with $\mathbb{E}(\gamma)= \bar\gamma$ being the average SNR.

\begin{table}[t]
\caption{Special cases in SOSF channel model}
\centering
  \begin{tabular}{ |l | c | c|c||c|c| }
	    \hline
    Fading & \textbf{$\omega_0$} & $\omega_1$ & $\omega_2$ & $\alpha$ & $\beta$  \\ \hline \hline
    Rice                 & - & - & 0 & 0 & -\\ \hline
		Double-Rayleigh \& LOS (DRLOS)        & - & 0 & - & \multicolumn{2}{c|}{$\alpha + \beta=1$} \\ \hline 
		Rayleigh and double-Rayleigh (RDR)  & 0 & - & - & - & 0 \\ \hline  
		Double-Rayleigh (DR)      & 0 & 0 & - & $1$ & 0 \\ \hline
		Rayleigh & 0 & - & 0 & 0 & 0\\ \hline		
		Static & - & 0 & 0 & 0 & 1\\ \hline
 \end{tabular}
\label{Tab:table_fading_types}
\end{table}

\section{Statistical analysis}
\label{Sec:Statistical analysis}
In this section, we provide new expressions for the SOSF statistics. Specifically, we will characterize the PDF, CDF and MGF of the receive SNR $\gamma$; note that the PDF and CDF expressions for the received signal envelope $|S|$ can be directly obtained by a simple change of variables.



\subsection{Derivation of the PDF}

Our goal is to obtain an expression for the PDF of the instantaneous receive SNR $\gamma$, $f_{\gamma}(\gamma)$. The starting point in our derivation is \eqref{Eq:Modelo_SOSF}, so that the squared signal envelope can be expressed as 

\begin {equation}
|S|^2=\frac{\gamma}{\bar\gamma}=|\omega_0 e^{j\phi}+\omega_1G_1+\omega_2 G_2 G_3|^2.
\label{Eq:Modelo_SOSF2}
\end{equation}

Since $G_3$ is a circularly-symmetric RV with zero-mean, we can express $G_3=|G_3|\cdot e^{j\phi_3}$, with $\phi_3$ uniformly distributed in $[0,2\pi)$ and $|G_3|$ being Rayleigh distributed. Because of $G_2$ being also circularly symmetric, the distribution of $G_2$ is the same as the distribution of $G_2\cdot e^{j\phi_3}$. Thus, we have that

\begin {equation}
|S|^2\buildrel d \over =|\omega_0 e^{j\phi}+\omega_1G_1+\omega_2 G_2  |G_3||^2=C.
\label{Eq:Modelo_SOSF3}
\end{equation}

Let us define the ancillary RV $X$ as $X\equiv |G_3|^2$, which follows an exponential distribution. Conditioning $S$ in (\ref{Eq:Modelo_SOSF3}) on a particular value of $X=x$ we have 
\begin{equation}
C|(X=x)=|\omega_0 e^{j\phi}+\omega_1G_1+\omega_2 \sqrt{x} G_2|^2. 
\label{Eq:Modelo_SOSF4}
\end{equation}
The two last terms in \eqref{Eq:Modelo_SOSF4} correspond to the sum of two RVs distributed as $\mathcal{N}_c(0,\omega_1^2)$ and $\mathcal{N}_c(0,\omega_2^2 x)$, respectively. This is equivalent to one single RV distributed as $\mathcal{N}_c(0,\omega_1^2+\omega_2^2 x)$. Hence, we have that $C|(X=x)$ is built from the sum of a LOS and a Rayleigh component. With all these considerations, the distribution of $\gamma|(X=x)\triangleq \gamma_x=\bar\gamma \cdot C|(X=x)$ is a conditional Rician distribution with PDF given by \cite{Simon2005}

\begin{multline}
\label{Eq:PDF_Rice}
f_{Rice}(\gamma;K_x,\bar\gamma_x)=\tfrac{1+K_x}{\bar \gamma_x} e^{-K_x}e^{-\frac{(1+K_x)\gamma}{\bar \gamma_x}} \\ \times I_0\left(2\sqrt{\tfrac{K_x(1+K_x)\gamma}{\bar \gamma_x}}\right), \;\;\;\; \gamma \geq 0
\end{multline}
where $I_0(\cdot)$ denotes the modified Bessel function of the first kind and zero order, with
\begin{align}
\bar \gamma_x \triangleq \mathbb{E}(\gamma_x)&= \omega_0^2+\omega_1^2+\omega_2^2 x=\bar \gamma (1-\alpha(1-x)),\label{Eq:gamma_x}\\
K_x&=\frac{\omega_0^2}{\omega_1^2+\omega_2^2 x}=\frac{\beta}{1-\beta-\alpha(1-x)}.\label{Eq:K_x}
\end{align}
Finally, averaging \eqref{Eq:PDF_Rice} over the distribution of $X$, we can obtain the PDF of the SOSF channel model as
\begin{equation}
\label{Eq:PDF_Rice_integrar}
f_{\gamma}(\gamma)=\int_{0}^{\infty} f_{Rice}(\gamma;K_x,\bar\gamma_x) \cdot f_{X}(x) dx,
\end{equation}
where $f_{X}(x)=e^{-x},\;\;x \geq 0$. Substituting (\ref{Eq:alpha_beta}), (\ref{Eq:gamma_x}) and (\ref{Eq:K_x}) into (\ref{Eq:PDF_Rice}) and (\ref{Eq:PDF_Rice_integrar}), we finally get

\begin{multline}
\label{Eq:PDF_SOSF}
f_{\gamma}(\gamma)=\int_{0}^{\infty}\frac{1}{(1-\beta-\alpha(1-x))\bar\gamma} e^{-\frac{\gamma+\beta\bar\gamma}{(1-\beta-\alpha(1-x))\bar\gamma}} \\ \times I_0\left(\frac{2}{(1-\beta-\alpha(1-x))}\sqrt{\frac{\beta \gamma}{\bar\gamma}} \right)e^{-x}dx.
\end{multline}

Equation (\ref{Eq:PDF_SOSF}) constitutes a novel expression for the PDF of the instantaneous SNR in a SOSF fading channel, and is an alternative to the one proposed in the literature \cite{Salo2006,Bandemer2009} that requires the integration of a highly oscillatory function over an infinite range, as:
\begin{equation}
\label{Eq:PDF_original}
f_{\gamma}(\gamma)=\int_{0}^{\infty}\frac{2J_0\left(\sqrt{\gamma}z\right)J_0\left(\sqrt{\beta \bar \gamma}z\right)}{4+\alpha\bar \gamma z^2}e^{-\frac{\bar \gamma (1-\beta-\alpha)z^2}{4}} z dz.
\end{equation}

Compared to the original expression in (\ref{Eq:PDF_original}), our approach has numerous benefits: first, the numerical integration in \eqref{Eq:PDF_SOSF} is well-behaved, as the integrand is the Rician PDF itself, instead of the product of two Bessel functions of the first kind and zero order $J_0(\cdot)$ that compromise conventional numerical methods. Secondly, as we will later see, expressing the SOSF channel as a continuous mixture of Rician channels greatly simplifies the use of this fading model for performance analysis purposes. Third, when $\omega_0=0$ (i.e. $\beta=0$) the special cases of the SOSF channel arise as a continuous mixture of Rayleigh fading channels, since $K_x=0$. This will also help simplifying the analysis of still rather complicated distributions such as the RDR case. Fourth, a direct observation of \eqref{Eq:Modelo_SOSF3} indicates that $S$ can be generated from only \emph{two} complex Gaussian RVs, instead of three as suggested by \eqref{Eq:Modelo_SOSF2}.

Specializing $\alpha$ and $\beta$ in (\ref{Eq:PDF_SOSF}) with the values indicated in Table \ref{Tab:table_fading_types} we can obtain a PDF expression for all particular fading distributions in closed-form, as summarized in Table \ref{Tab:PDF}. For the Rician, Rayleigh and static cases, the derivation is straightforward and the result is well-known, so that they are omitted in the Table for the sake of compactness. For the other three cases we have:

\subsubsection{DRLOS} Setting $\alpha+\beta=1$ in (\ref{Eq:PDF_SOSF}), the integral reduces to that in \cite[eq. (6.653)]{gradshteyn2014}, which matches the result proposed in \cite{Vinogradov2015} as a piecewise function shown in Table \ref{Tab:PDF}, equation (\ref{Eq:PDF_DRLOS}).
\subsubsection{RDR} Setting $\beta=0$ in (\ref{Eq:PDF_SOSF}), the integral matches the definition of the generalized incomplete gamma function\footnote{Not to be confused with a different generalization of the incomplete gamma function defined as $\Gamma(a,z_1,z_2)=\int_{z_1}^{z_2}t^{a-1}e^{-t}dt$.} in \cite{CHAUDHRY199499} and also used in \cite{Gaaloul2012,Siddiqui2018}, as $\Gamma(a,x,b)=\int_{x}^{\infty}t^{a-1} e^{-t}e^{\frac{-b}{t}}dt$ yielding the RDR PDF shown in Table \ref{Tab:PDF}, equation (\ref{Eq:PDF_RDR}). 
\subsubsection{DR} Setting $\beta=0$ and $\alpha=1$ in (\ref{Eq:PDF_DRLOS}) we get the DR distribution PDF shown in table \ref{Tab:PDF}, equation (\ref{Eq:PDF_DR}) which is in agreement with the expression proposed in \cite{Kovac2002}.

\begin{table*}[t]
\caption{SOSF channel PDFs (special cases).}
\centering
  \begin{tabular}{ |l|c|}
	  \hline
    \textbf{Fading} & $f_{\gamma}(\gamma),\;\;\;\gamma>0$  \\  \hline
    SOSF                 &  \parbox{10cm}{\begin {equation}\int_{0}^{\infty}\frac{1}{(1-\beta-\alpha(1-x))\bar\gamma} e^{-\frac{\gamma+\beta\bar\gamma}{(1-\beta-\alpha(1-x))\bar\gamma}}  I_0\left(\frac{2}{(1-\beta-\alpha(1-x))}\sqrt{\frac{\beta \gamma}{\bar\gamma}} \right)e^{-x}dx \tag{\ref{Eq:PDF_SOSF}}\end{equation}}\\ \hline 
		DRLOS       & \parbox{10cm}{\begin {equation}\label{Eq:PDF_DRLOS}
    \begin{cases}
      \frac{2}{(1-\beta)\bar\gamma}I_0\left(\frac{2\sqrt{\gamma}}{\sqrt{(1-\beta)\bar\gamma}}\right)K_0\left(2\sqrt{\frac{\beta}{{1-\beta}}}\right), &0<\gamma<\beta\bar\gamma \\
			\frac{2}{(1-\beta)\bar\gamma}K_0\left(\frac{2\sqrt{\gamma}}{\sqrt{(1-\beta)\bar\gamma}}\right)I_0\left(2\sqrt{\frac{\beta}{{1-\beta}}}\right), &\gamma>\beta\bar\gamma\\
			     \end{cases}
\end{equation}}\\ 
		 \hline 
		RDR  & \parbox{11cm}{\begin {equation}\label{Eq:PDF_RDR}\frac{1}{\alpha \bar \gamma}e^{\frac{1-\alpha}{\alpha}} \Gamma\left(0,\frac{1-\alpha}{\alpha},\frac{\gamma}{\alpha \bar \gamma} \right)\end{equation}}\\ \hline
		DR      & \parbox{11cm}{\begin {equation}\label{Eq:PDF_DR} \frac{2}{\bar\gamma}K_0\left(\frac{2\sqrt{\gamma}}{\sqrt{\bar\gamma}}\right)\end{equation}}\\ \hline
	 \end{tabular}
\label{Tab:PDF}
\end{table*}


\subsection{Derivation of the CDF}
Following the same procedure as with the PDF, we can obtain an expression for the CDF of the SOSF model by averaging the CDF of a Rician distribution conditioned to a fixed value $X=x$. The CDF of $\gamma_x$ is given in \cite{Simon2005} as

\begin{equation}
\label{Eq:CDF_Rice}
F_{Rice}(\gamma;K_x,\bar\gamma_x)=1-Q_1\left(\sqrt{2K_x},\sqrt{\frac{2(1+K_x)\gamma}{\bar\gamma_x}}\right),
\end{equation}

where $Q_1(\cdot,\cdot)$ is the Marcum Q-function and $\bar \gamma_x$ and $K_x$ are those previously defined in (\ref{Eq:gamma_x}) and (\ref{Eq:K_x}). Averaging (\ref{Eq:CDF_Rice}) over the values of $X$, we obtain
\begin{multline}
\label{Eq:CDF_SOSF}
F_{\gamma}(\gamma)=\\ 1-\int_{0}^{\infty}Q_1\left(\sqrt{\tfrac{2 \beta}{1-\beta-\alpha(1-x)}},\sqrt{\tfrac{2\gamma /\bar\gamma}{1-\beta-\alpha(1-x)}}\right)e^{-x}dx.
\end{multline}

Again, equation (\ref{Eq:CDF_SOSF}) is a novel expression for the CDF of the instantaneous SNR in a SOSF fading channel. Because the integrand is a CDF itself, it does not oscillate and is a numerically well-conditioned alternative to the expression commonly used in the literature \cite{Salo2006,Bandemer2009}. 

The CDF of the special cases of the SOSF channel can be obtained either by specializing $\alpha$ and $\beta$ in (\ref{Eq:CDF_SOSF}) according to Table \ref{Tab:table_fading_types}, or by integrating the corresponding PDFs shown in Table \ref{Tab:PDF}. This second approach yields closed-form solutions in a rather straightforward procedure as follows:
\subsubsection{DRLOS} The integration of (\ref{Eq:PDF_DRLOS}) with respect to $\gamma$ yields an integral that is solved in \cite[eq. (6.561.7) and (6.561.8)]{gradshteyn2014}, resulting in the closed-form CDF expression in Table \ref{Tab:CDF}, equation (\ref{Eq:CDF_DRLOS})
\subsubsection{RDR} In order to integrate (\ref{Eq:PDF_RDR}) with respect to $\gamma$, we use the relationship $\int_{0}^{b}\Gamma(0,x,z)dz=e^{-x}-\Gamma(1,x,b)$ (see Appendix \ref{app1}) to obtain the expression for the RDR CDF given in table \ref{Tab:CDF}, equation (\ref{Eq:CDF_RDR}).
\subsubsection{DR} Setting $\beta=0$ and $\alpha=1$ in (\ref{Eq:CDF_DRLOS}), we get the DR distribution CDF shown in table \ref{Tab:CDF}, equation (\ref{Eq:CDF_DR}). The obtained expression is in accordance to that proposed in \cite{Kovac2002}.

\begin{table*}[t]
\caption{SOSF channel CDFs (special cases).}
\centering
 \begin{tabular}{ |l|c|}
    \hline
    \textbf{Fading} & $F_{\gamma}(\gamma) \;\;\; \gamma>0$ \\ \hline \hline
    SOSF                 & \parbox{11cm}{\begin {equation}1-\int_{0}^{\infty}Q_1\left(\sqrt{\tfrac{2 \beta}{1-\beta-\alpha(1-x)}},\sqrt{\tfrac{2\gamma /\bar\gamma}{(1-\beta-\alpha(1-x))}}\right)e^{-x}dx\tag{\ref{Eq:CDF_SOSF}}\end{equation}}\\ \hline  \hline
		DRLOS       &  \parbox{11cm}{\begin {equation}\label{Eq:CDF_DRLOS}
    \begin{cases}
      \frac{2}{\sqrt{(1-\beta)}}K_0\left(2\sqrt{\tfrac{\beta}{{1-\beta}}}\right)\sqrt{\frac{\gamma}{\bar \gamma}}I_1\left( 2\sqrt{\tfrac{\gamma/\bar\gamma}{(1-\beta)}}\right),\;\;0<\gamma<\beta\bar\gamma \\
			c+\frac{2}{\sqrt{(1-\beta)}}I_0\left(2\sqrt{\tfrac{\beta}{1-\beta}} \right) \left[ \sqrt{\beta}K_1\left( 2\sqrt{\tfrac{\beta}{(1-\beta)}}\right)-\sqrt{\frac{\gamma}{\bar \gamma}}K_1\left(2\sqrt{\tfrac{\gamma/\bar\gamma}{(1-\beta)}}\right)\right],\;\;\gamma>\beta\bar\gamma\\
     \end{cases}
\end{equation}}\\
	&	\text{with} {$c=2\sqrt{\tfrac{\beta}{(1-\beta)}}K_0\left(2\sqrt{\tfrac{\beta}{{1-\beta}}}\right)I_1\left(2\sqrt{\tfrac{\beta}{{1-\beta}}}\right)$} 
\\ \hline
		RDR  & \parbox{11cm}{\begin {equation}\label{Eq:CDF_RDR} 1-e^{\frac{1-\alpha}{\alpha}} \Gamma\left(1,\frac{1-\alpha}{\alpha},\frac{\gamma}{\alpha \bar \gamma} \right)\end{equation}} \\ \hline 
		DR    &  \parbox{11cm}{\begin {equation}\label{Eq:CDF_DR} 1-2\sqrt{\tfrac{\gamma}{\bar \gamma}}K_1\left(2\sqrt{\tfrac{\gamma}{\bar \gamma}} \right)\end{equation}}\\ \hline
   \end{tabular}
\label{Tab:CDF}
\end{table*}


\subsection{Derivation of the MGF}
The MGF of the SNR for the SOSF channel model is not known. A direct inspection of \eqref{Eq:Modelo_SOSF4} reveals that the MGF of the conditional SNR $\gamma_x$ is that of the Rician distribution, as \cite{Simon2005}

\begin {equation}
\label{Eq:Rice_MGF}
\mathcal{M}_{Rice}(s;K_x,\bar\gamma_x)=\frac{1+K_x}{1+K_x-s\bar\gamma_x}e^{\frac{K_x\bar \gamma_x s}{K+1-s\bar\gamma_x}},
\end {equation}
hence
\begin {equation}
\label{Eq:SOSF_Conditioned_MGF}
\mathcal{M}_{\gamma}(s)=\int_{0}^{\infty}\mathcal{M}_{Rice}(s;K_x,\bar\gamma_x)\cdot f_{X}(x)dx.
\end {equation}

Substituting $K_x$ and $\gamma_x$ from (\ref{Eq:K_x}) and (\ref{Eq:gamma_x}) in (\ref{Eq:Rice_MGF}), and then (\ref{Eq:Rice_MGF}) in (\ref{Eq:SOSF_Conditioned_MGF}), we can write after some manipulations
\begin {equation}
\label{Eq:SOSF_Conditioned_MGF_II}
\mathcal{M}_{\gamma}(s)=\frac{-e^{\frac{s\bar \gamma(1-\alpha-\beta)-1}{s\bar\gamma \alpha}}}{s\bar \gamma \alpha} \int_{\frac{1-s(1-\alpha-\beta)\bar\gamma}{P-s\alpha \bar \gamma}}^{\infty} \frac{1}{z}\;e^{-\frac{\beta}{\alpha z}}\;e^{-z}dz.
\end{equation}

The integral in (\ref{Eq:SOSF_Conditioned_MGF_II}) corresponds to the previously introduced generalized incomplete gamma function \cite{CHAUDHRY199499}, so we can finally write

\begin {equation}
\label{Eq:MGF_SOSF}
\mathcal{M}_{\gamma}(s)=\frac{-e^{\frac{s\bar \gamma(1-\alpha-\beta)-1}{s\bar\gamma \alpha}}}{s\bar \gamma \alpha} \Gamma \left(0,\frac{s(1-\alpha-\beta)\bar \gamma-1}{s\alpha\bar \gamma},\frac{\beta}{\alpha}\right).
\end{equation}

Expression \eqref{Eq:MGF_SOSF} is new in the literature to the best of our knowledge. Specializing the values of $\alpha$ and $\beta$ in (\ref{Eq:MGF_SOSF}) yields the MGF expressions for the special cases of the SOSF distribution summarized in Table \ref{Tab:MGF}, which are derived as follows:

\subsubsection{DRLOS} Setting $\alpha+\beta=1$ in (\ref{Eq:MGF_SOSF}) results in expression (\ref{Eq:MGF_DRLOS}) in Table \ref{Tab:MGF}.
\subsubsection{RDR} Setting $\beta=0$ in (\ref{Eq:MGF_SOSF}), and using the same identity as in the DR case we obtain the expression (\ref{Eq:MGF_RDR}) given in Table \ref{Tab:MGF}.
\subsubsection{DR} Setting $\beta=0$ and $\alpha=1$ in (\ref{Eq:MGF_SOSF}), and using the identity $\Gamma(0,x,0)=E_1(x)$ where $E_1(\cdot)$ is the exponential integral defined as $E_1(x)=\int_{x}^{\infty}\frac{e^{-z}}{z}dz$, we get the MGF of the DR case in Table \ref{Tab:MGF}, equation (\ref{Eq:MGF_DR}).

\begin{table}[t]
\caption{SOSF channel MGFs (special cases).}
\centering
\begin{tabular}{ |l|c|}
    \hline
    \textbf{Fading} & $MGF_{\gamma}(s) \;\;\; s<0$ \\ \hline \hline
    SOSF                 & \parbox{7cm}{\begin{equation}\frac{-e^{\frac{s\bar \gamma(1-\alpha-\beta)-1}{s\bar\gamma \alpha}}}{s\bar \gamma \alpha} \Gamma \left(0,\frac{s(1-\alpha-\beta)\bar \gamma-1}{s\alpha\bar \gamma},\frac{\beta}{\alpha}\right)\tag{\ref{Eq:MGF_SOSF}}\end{equation}}\\ \hline  
		DRLOS        & \parbox{7cm}{\begin{equation} \label{Eq:MGF_DRLOS}\frac{-e^{\frac{-1}{s\bar\gamma \alpha}}}{s\bar \gamma \alpha} \Gamma \left(0,\frac{-1}{s\alpha\bar \gamma},\frac{1-\alpha}{\alpha}\right) \end{equation}}\\ \hline
		RDR  & \parbox{7cm}{\begin{equation} \label{Eq:MGF_RDR}-\frac{e^{\frac{s\bar \gamma(1-\alpha)-1}{s\bar \gamma \alpha}}}{s\bar \gamma \alpha}E_1\left( \tfrac{s(1-\alpha)\bar \gamma-1}{s\bar \gamma \alpha}\right)\end{equation}} \\ \hline  
		DR      & \parbox{7cm}{\begin{equation} \label{Eq:MGF_DR}-\frac{e^{-\frac{1}{s\bar \gamma}}}{s\bar \gamma}E_1\left( \tfrac{-1}{s\bar \gamma}\right)\end{equation}}\\ \hline
 \end{tabular}
\label{Tab:MGF}
\end{table}

\subsection{Tail approximation for the CDF}
It is also possible to obtain a tail approximation for the SOSF CDF in the form of
\begin{equation}
\label{eqnew01}
F_{\gamma}(\gamma)\approx \frac{a}{d}\left(\frac{\gamma}{\bar\gamma}\right)^{d},
\end{equation}
where $d$ is the diversity order and $a$ is a power offset also related to the coding gain \cite{Wang2003}.

Conditioning to $(X=x)$, the tail approximation of the conditioned SOSF distribution is given by
\begin{equation}
\label{Eq:CDF_tail_generic}
F_{\gamma}(\gamma) \approx \frac{a_x}{d}\left( \frac{\gamma}{\bar \gamma_x}\right)^d,
\end{equation}
where $d=1$ and $a_x=(1+K_x)e^{-K_x}$ \cite{Wang2003}, with $K_x$ and $\bar \gamma_x$ given by (\ref{Eq:gamma_x}) and (\ref{Eq:K_x}). Integrating (\ref{Eq:CDF_tail_generic}) over $x$, we obtain the desired tail approximation for the SOSF distribution in \eqref{eqnew01}, with $d=1$ and 

\begin{equation}
\label{Eq:CDF_tail_a_SOSF}
a=\frac{e^{\frac{1-\beta-\alpha}{\alpha}}}{\alpha} \Gamma \left(0,\tfrac{1-\beta-\alpha}{\alpha},\tfrac{\beta}{\alpha}  \right).
\end{equation}

\section{Performance analysis over SOSF channels}
\label{Application example: Ergodic Capacity}
\subsection{Formulation}

One of the key observations of our approach is the fact that the SOSF channel can be regarded as a continuous mixture of Rician fading channels. This somehow resembles similar connections recently unveiled in the literature, which express the two-wave with diffuse power (TWDP) fading channel as a continuous mixture of Rician fading channels \cite{Rao2015}, or the Nakagami-$q$ fading channel as a continuous mixture of Rayleigh fading channels \cite{Romero2017b}. Leveraging this fundamental connection between the SOSF distribution and an underlying conditional Rician distribution, it is possible to analyze general performance metrics for the SOSF channel by using readily available results for the Rician case. This is formally stated in the following lemma, as follows:

\begin{lemma} \label{l2}
Let $H(\gamma)$ be any arbitrary performance metric depending on the instantaneous SNR $\gamma$, and let $\overline{H}_R(\overline{\gamma},K)$ be such performance metric in Rician fading with average SNR $\overline{\gamma}$ and Rician parameter $K$, obtained by averaging over an interval of the PDF of the SNR, i.e.,
\begin{equation} \label{eq:15}
	 \overline{H}_R(\overline{\gamma},K)= \int_a^b 
	 H(y) f_{\gamma}\left(y;\bar\gamma,K\right)
	 dy,
\end{equation}
with $0 \leq a < b \leq \infty$, and $f_{\gamma}\left(\gamma;\bar\gamma,K\right)$ is the PDF of the SNR in Rician fading. Then, the performance metric in SOSF channels with average SNR $\overline{\gamma}$ and shape parameters $\alpha$ and $\beta$, denoted as $\overline{H}_S(\overline{\gamma};\alpha,\beta)$, can be calculated as
\begin{equation} \label{eq:16}
	 \overline{H}_S(\overline{\gamma};\alpha,\beta)= \int_0^\infty 
	\overline{H}_R(\overline{\gamma_x},K_x)e^{-x}dx,
\end{equation}
where $\overline{\gamma_x}$ and $K_x$ are defined in \eqref{Eq:gamma_x} and \eqref{Eq:K_x}.
\end{lemma}
\begin{IEEEproof}
The performance metric $\overline{H}_S(\overline{\gamma};\alpha,\beta)$ is obtained as
\begin{equation} \label{eq:16_1}
	 \overline{H}_S(\overline{\gamma};\alpha,\beta)= \int_a^b 
	 H(y) f_{\gamma}(y;\alpha,\beta)
	 dy,
\end{equation}
where $f_{\gamma}(x;\alpha,\beta)$ is the PDF of the SNR in SOSF channels given in \eqref{Eq:PDF_SOSF}, where the dependence on $\alpha$ and $\beta$ is indicated for the sake of notational clarity.
Thus, we can write
\begin{align} \label{eq:17}
\overline{H}_S&(\overline{\gamma};\alpha,\beta)=\int_a^b H(y) \left\{\int_0^\infty 
	 \frac{ e^{-\frac{y+\beta\bar\gamma}{(1-\beta-\alpha(1-x))\bar\gamma}}}{(1-\beta-\alpha(1-x))\bar\gamma}\right.\nonumber \\ &\left. \times I_0\left(\frac{2}{(1-\beta-\alpha(1-x))}\sqrt{\frac{\beta y}{\bar\gamma}} \right)e^{-x}  dx\right\}
	 dy.
\end{align}
By reversing the order of integration\footnote{As stated in \cite{Romero2017}, a sufficient condition for this double integral to be reversible is that $H(x)$ is a nonnegative continuous function. This is the case of conventional performance metrics of interest, such as channel capacity, symbol error rate or outage probability.
}
and using \eqref{Eq:PDF_Rice}, we have
\begin{align} \label{eq:18}
\overline{H}_S&(\overline{\gamma};\alpha,\beta)= \int_0^\infty \left\{\int_a^b H(y)
	 \frac{1+K_x}{\bar \gamma_x} e^{-K_x}e^{-\frac{(1+K_x)y}{\bar \gamma_x}}\right.\nonumber \\ &\left. \times I_0\left(2\sqrt{\tfrac{K_x(1+K_x)y}{\bar \gamma_x}}\right) dy \right\}e^{-x}dx,
\end{align}
where $\overline{\gamma_x}$ and $K_x$ are given in \eqref{Eq:gamma_x} and \eqref{Eq:K_x}. Identifying the inner integral as $\overline{H}_R(\gamma_x,K_x)$, (\ref{eq:16}) is finally obtained.  
\end{IEEEproof}

Lemma \ref{l2} can be seen as a general framework to analyze the performance of communication systems operating in SOSF channels, for which previous results are available for the Rician case. We will exemplify its use in the following subsection, by conducting a capacity analysis in SOSF channels.

%


\subsection{Application example: average capacity}
The average channel capacity in fading channels is defined as
\begin{equation}
\label{Eq:Ergodic_Capacity_def}
C=B \int_{0}^{\infty}\log_2(1+\gamma)f_{\gamma}(\gamma) d\gamma,
\end{equation}
where $B$ is the bandwidth, $\gamma$ is the instantaneous receive SNR and $f_{\gamma}(\gamma)$ is the PDF of $\gamma$, i.e. the fading distribution. This capacity concides with the ergodic capacity of a fading channel with an optimal rate adaptation policy and constant transmit power, for which channel state information is only available at the receiver side \cite[eq. (8)]{Goldsmith1997}. Without loss of generality, we will consider a normalized bandwith $B=1$ in the following derivations.

Even though the average capacity is known for most popular fading channels distributions (see \cite{Laureano2016} and the references therein), this is not the case for the SOSF channel. Substituting $f_{\gamma}(\gamma)$ by (\ref{Eq:PDF_SOSF}) in (\ref{Eq:Ergodic_Capacity_def}), the capacity is obtained in a double integral form. However, because of the appealing numerical properties of the PDF in (\ref{Eq:PDF_SOSF}), such integral can be efficiently evaluated numerically.

In order to obtain a better insight into the effect of the SOSF channel parameters $\alpha$ and $\beta$ on the channel capacity, we carry out an asymptotic analysis in the high-SNR regime using the approach in \cite[eq. (8) and (9)]{Alouini2012}, which yields a tight lower bound for the capacity. Starting from the asymptotic capacity in the high-SNR regime for the Rician case in \cite[eq. (64)]{Rao2015}
\begin {equation}
\label{Eq:C_Rice}
\left. C_{Rice} \right|_{\bar \gamma \Uparrow} \approx \log_2(\bar \gamma)+\log_2\left(\tfrac{K}{1+K}\right)+\log_2(e)E_1(K),
\end{equation}
a direct application of Lemma \ref{l2} yields the asymptotic capacity for the SOSF fading channel as

\begin {align}
\label{Eq:C_asint_SOSF}
& \left. C\right|_{\bar \gamma \Uparrow} \approx log_2(\bar \gamma)-t,\\
t\triangleq &-log_2(\beta)-log_2(e)\int_{0}^{\infty}E_1\left(\tfrac{\beta}{1-\beta-\alpha(1-x)}\right) e^{-x}dx,
\end {align}
where the parameter $t$ can be regarded as a capacity loss with respect to the AWGN case. This expression is new in the literature to the best of our knowledge. We note that the integral term in \eqref{Eq:C_asint_SOSF} only needs to be evaluated once, as it does not depend on the average SNR $\bar\gamma$. For all special cases of the SOSF channel, it is possible to obtain a closed-form expression for the asymptotic capacity; these are summarized in Table \ref{Tab:Capacity}, where $\gamma_e$ is the Euler-Mascheroni constant. Specifically, expressions for DRLOS, DR and RDR are obtained as follows:

\subsubsection{DRLOS} Setting $\alpha+\beta=1$ in (\ref{Eq:C_asint_SOSF}), the integral can be solved using \cite[eq. (6.226) ]{gradshteyn2014}. This yields expression (\ref{Eq:C_DRLOS}) of Table \ref{Tab:Capacity}.
\subsubsection{DR} We must first set $\alpha=1$, and take the limit of (\ref{Eq:C_asint_SOSF}) when $\beta \rightarrow 0$. Taking into account that $E_1(x)\rightarrow -\gamma_e-\log(x)$ when $x \rightarrow 0$, and using \cite[eq. (4.331.1)]{gradshteyn2014}, expression (\ref{Eq:C_DR}) in Table \ref{Tab:Capacity} is obtained after some algebraic manipulation. The resulting expression agrees with that proposed in \cite{Shin2004}.
\subsubsection{RDR} Proceeding in the same manner as with the DR case, and using the identity \cite[eq. (4.337.1) ]{gradshteyn2014}, we reach the expression shown in equation (\ref{Eq:C_RDR}) of Table \ref{Tab:Capacity}.

\begin{table}[t]
\caption{Asymptotic average capacity of SOSF channel in the high-SNR regime}
\centering
    \begin{tabular}{ |l|c|}
    \hline
    \textbf{Fading} & Asymptotic capacity \\ \hline \hline
    SOSF                 & \parbox{7cm}{\begin{equation}log_2(\bar \gamma)+log_2(\beta)+log_2(e)\int_{0}^{\infty}E_1\left(\tfrac{\beta}{1-\beta-\alpha(1-x)}\right) e^{-x}dx \tag{\ref{Eq:C_asint_SOSF}}\end{equation}}\\ \hline  
		Rice                 & \parbox{7cm}{\begin{equation}\label{Eq:C_Rice}log_2(\bar \gamma)+log_2(\beta)+log_2(e)E_1\left(\tfrac{\beta}{1-\beta}\right)\end{equation}} \\ \hline
		DRLOS        & \parbox{7cm}{\begin{equation}\label{Eq:C_DRLOS}log_2(\bar \gamma)+log_2(\beta)+log_2(e)2K_0\left(2\sqrt{\tfrac{\beta}{1-\beta}}\right)\end{equation}}\\ \hline
		RDR  & \parbox{7cm}{\begin{multline}\label{Eq:C_RDR}log_2(\bar \gamma)-log_2(e)\gamma_e+ \\log_2(1-\alpha)+log_2(e)e^{\tfrac{1-\alpha}{\alpha}}E_1\left( \frac{1-\alpha}{\alpha}\right)\end{multline}}\\ \hline 
		Rayleigh & \parbox{7cm}{\begin{equation}\label{Eq:C_R}log_2(\bar \gamma)-log_2(e)\gamma_e\end{equation}} \\\hline
		DR     & \parbox{7cm}{\begin{equation}\label{Eq:C_DR}log_2(\bar \gamma)-2log_2(e)\gamma_e\end{equation}}\\ \hline
		AWGN &\parbox{7cm}{\begin{equation}\label{Eq:C_AWGN}log_2(\bar \gamma)\end{equation}} \\ \hline
   \end{tabular}
	\label{Tab:Capacity}
	\end{table}
	
\subsection{Numerical results}

\begin{figure}[t]
\centering
\includegraphics[width=.5\textwidth]{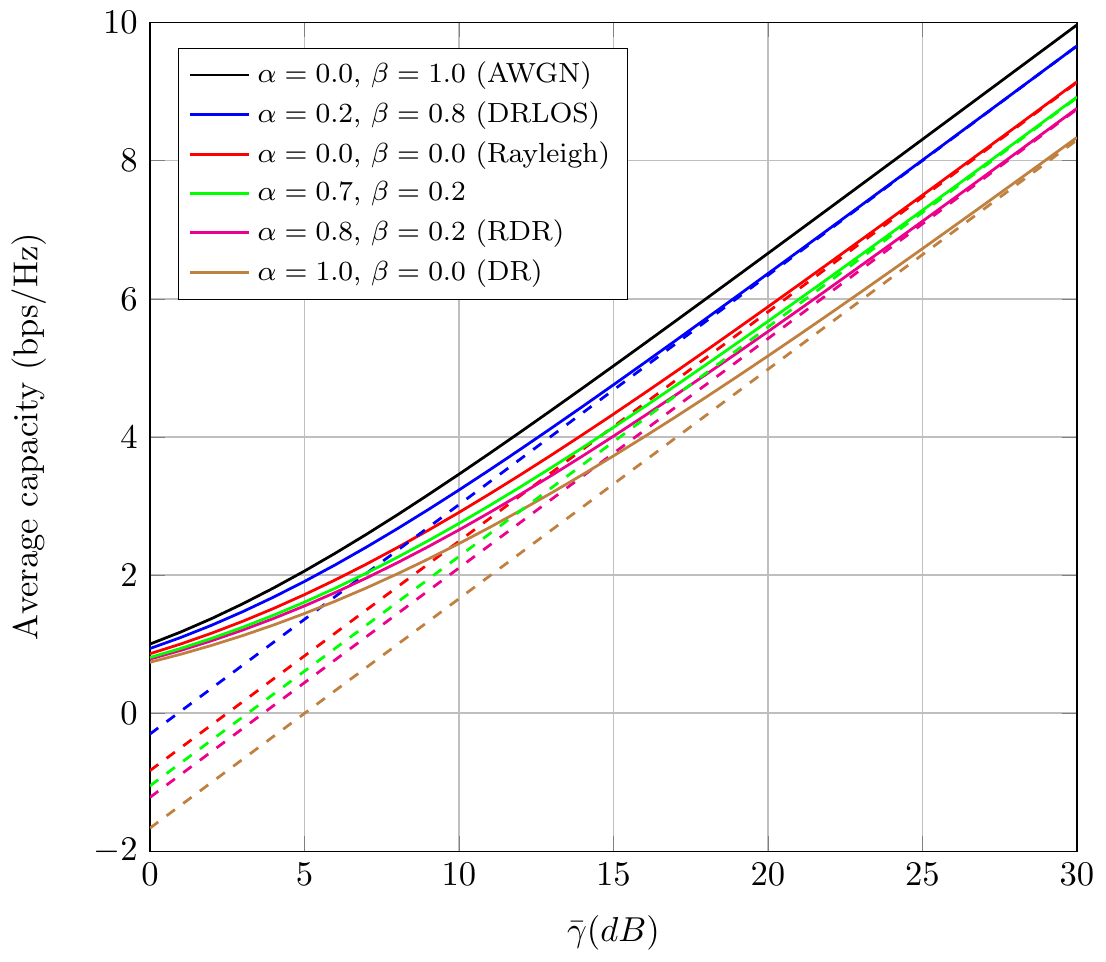}
\centering
\caption{Exact (solid line) and asymptotic (dashed line) average capacity of different SOSF channels.} 
\label{Fig_Capa_gamma}
\end{figure}

In Fig. \ref{Fig_Capa_gamma}, the exact and asymptotic capacity of different SOSF channels using the expressions previously derived are depicted as a function of the average SNR $\bar \gamma$. The AWGN channel capacity (corresponding to the static case) is also included as an upper bound, which will be used as a benchmark for comparison purposes. The exact capacity has been calculated by substituting (\ref{Eq:PDF_SOSF}) in (\ref{Eq:Ergodic_Capacity_def}) and performing numerical integration, while the asymptotic capacity has been obtained using the corresponding expressions shown in Table \ref{Tab:Capacity}. Results have also been validated with Monte Carlo simulations, which are not overimposed in the figure for the sake of clarity. As expected, the asymptotic capacity tends to the exact capacity for high SNR for all the cases considered in the plot. We also observe a worse-than-Rayleigh behavior for some of the SOSF combinations due to the presence of the DR component. 

\begin{figure}[t]
\centering
\includegraphics[width=.99\columnwidth]{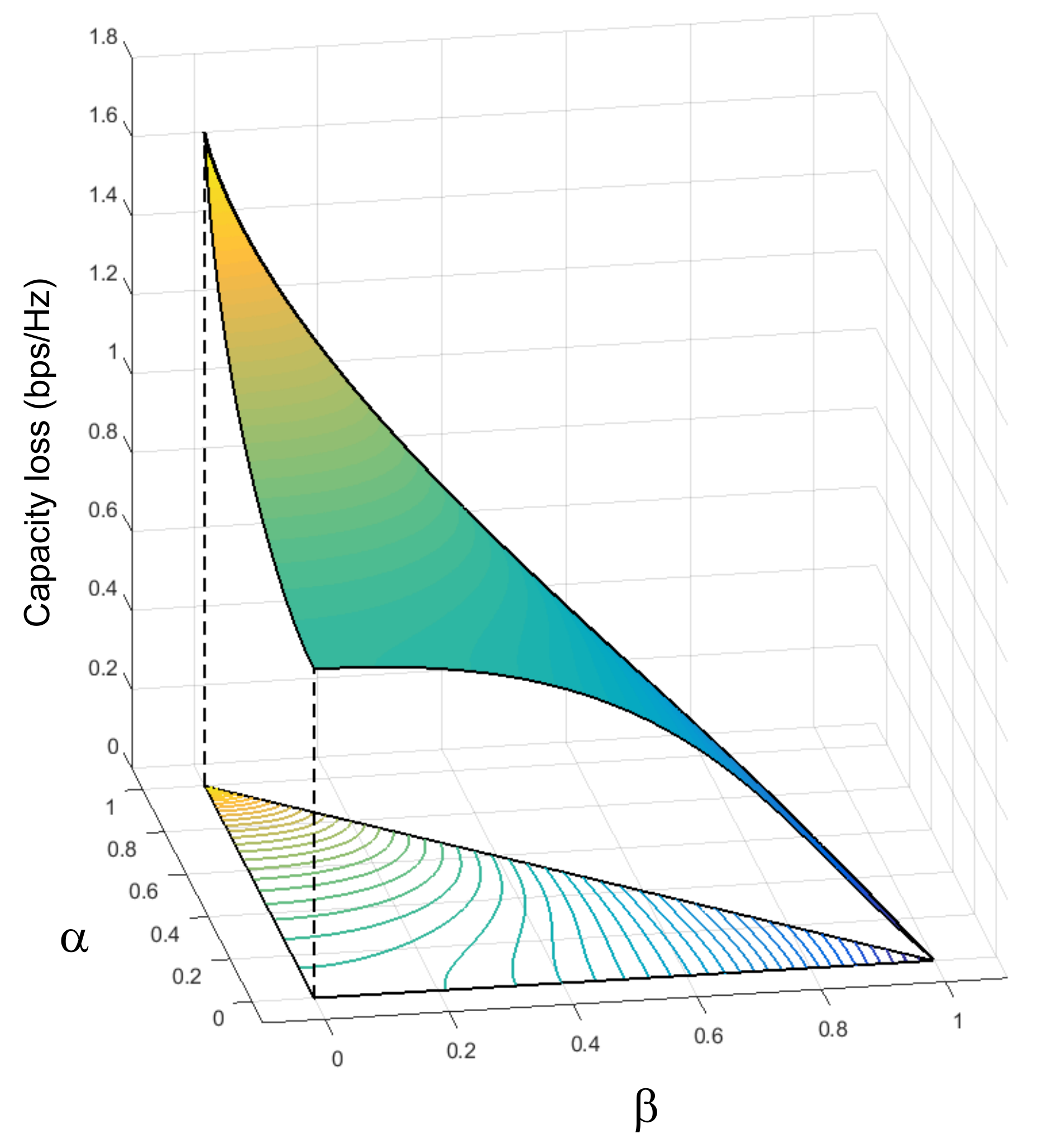}
\centering
\caption{Average capacity loss $t$ as a function of $\alpha$ and $\beta$, for $\bar \gamma=$40dB} 
\label{Fig_C_3D}
\end{figure}

The dependence of the asymptotic capacity on the SOSF shape parameters $\alpha$ and $\beta$ is better visualized in the 3D plot in Fig. \ref{Fig_C_3D}, on which the capacity loss with respect to the AWGN case $t$ is plotted as a function of $\alpha$ and $\beta$, for all the values within the valid triangular domain (see Fig.\ref{Fig_Triangle}), considering a sufficiently high average SNR $\bar \gamma=40 dB$.

We see that the cut obtained by setting $\alpha=0$ in the 3D plot corresponds to a Rician distribution (\ref{Eq:C_Rice}) and runs from the Rayleigh at coordinate $(\alpha,\beta)=(0,0)$ with a capacity loss of $\log_2(e)\gamma_{e}=0.83$ bps/Hz to the AWGN case at coordinate $(0,1)$ which is the reference capacity value. Setting $\beta=0$, the resulting curve corresponds to the RDR distribution (\ref{Eq:C_RDR}) that ranges from the Rayleigh distribution at $(0,0)$ to the DR distribution at coordinate $(1,0)$. We have also highlighted the DRLOS curve (\ref{Eq:C_DRLOS}) obtained by setting $\alpha+\beta=1$, that starts in the AWGN case and ends in the DR distribution. Notice that the maximum capacity loss for the SOSF channel corresponds to the DR case, taking the value $2\log_2(e)\gamma_{e}=1.67$ bps/Hz, i.e. twice as much as in the Rayleigh case.

From the inspection of Fig. \ref{Fig_C_3D} we may conclude that in general terms, the asymptotic capacity loss decreases with increasing $\beta$ (LOS component) and decreasing $\alpha$ (DR component) which is a logical behavior. Although this is true for parameter $\beta$ (capacity loss decreases with increasing LOS component), a closer look into the variation of the asymptotic capacity loss with respect to $\alpha$ reveals a somehow unexpected behavior. For a fixed value of $\beta$, the value of $\alpha$ indicates the relative weight between the Rayleigh and the DR components, with $\alpha=0$ corresponding to only Rayleigh component and $\alpha=1-\beta$ to only DR component. Given that the DR component is more fluctuating than the Rayleigh one, an increase in the capacity loss (i.e. a worse capacity) as $\alpha$ grows would be expected. This can be seen for instance in the curve corresponding to the RDR distribution ($\beta=0$) in Fig.\ref{Fig_C_3D}. However, this is not true for the whole range of $\beta$. 

\begin{figure}[t]
\centering
\includegraphics[width=.99\columnwidth]{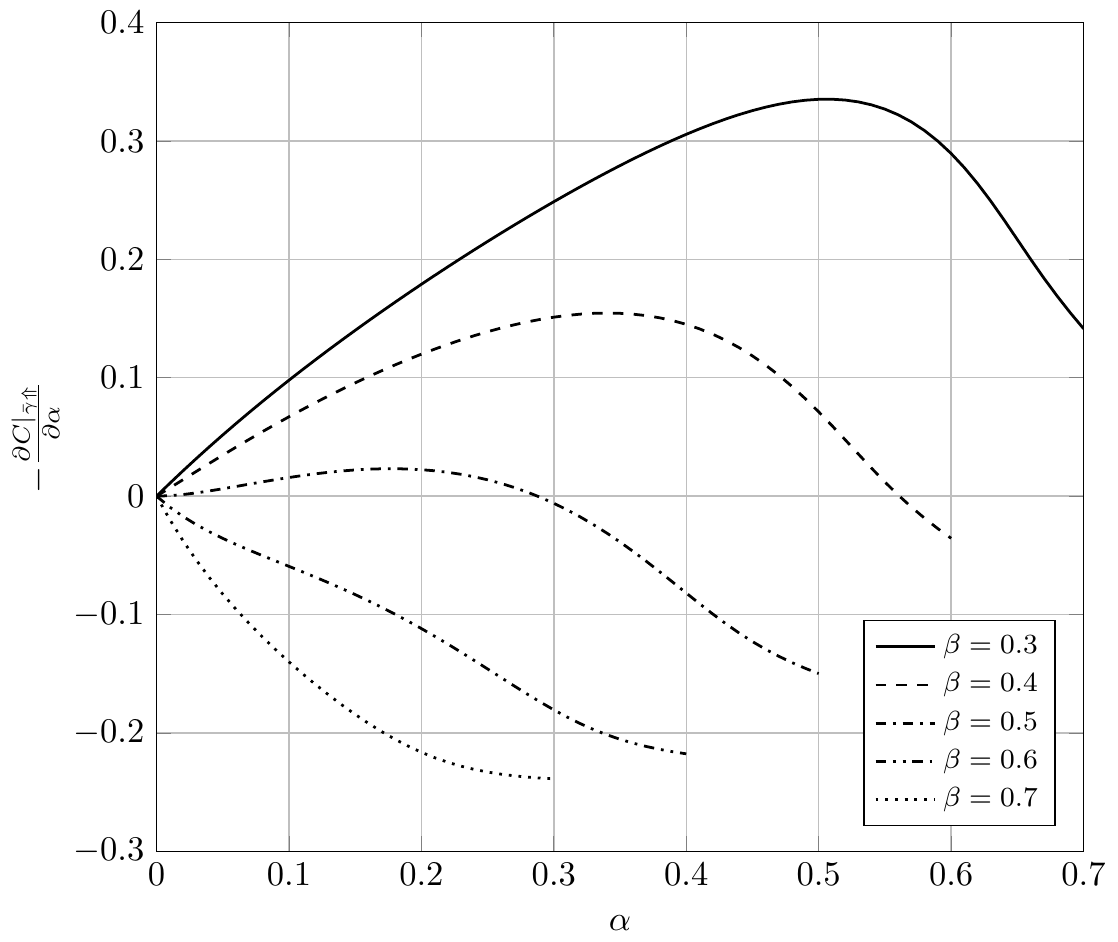}
\centering
\caption{Derivative of the average capacity loss with respect to $\alpha$ for different values of $\beta$} 
\label{Fig_C_derivada_alpha}
\end{figure}

In order to better illustrate this behavior, we plot the derivative of the asymptotic capacity loss with respect to $\alpha$ in Fig. \ref{Fig_C_derivada_alpha}, i.e.  $\tfrac{\partial \left( \log_2(\bar \gamma)- C|_{\bar \gamma \Uparrow}\right)}{\partial \alpha}=-\frac{\partial C|_{\bar \gamma \Uparrow}}{\partial \alpha}$, for different values of $\beta$ using (\ref{Eq:C_asint_SOSF}). We see that this derivative takes negative values, specially for high values of $\beta$ (i.e. stronger LOS). This implies a decrease in capacity loss i.e. a larger capacity as $\alpha$ grows. In other words, in the presence of a significant LOS component, the SOSF channel has a larger capacity with a higher proportion of a DR component than a Rayleigh one.

\begin{figure}[t]
\centering
\includegraphics[width=.99\columnwidth]{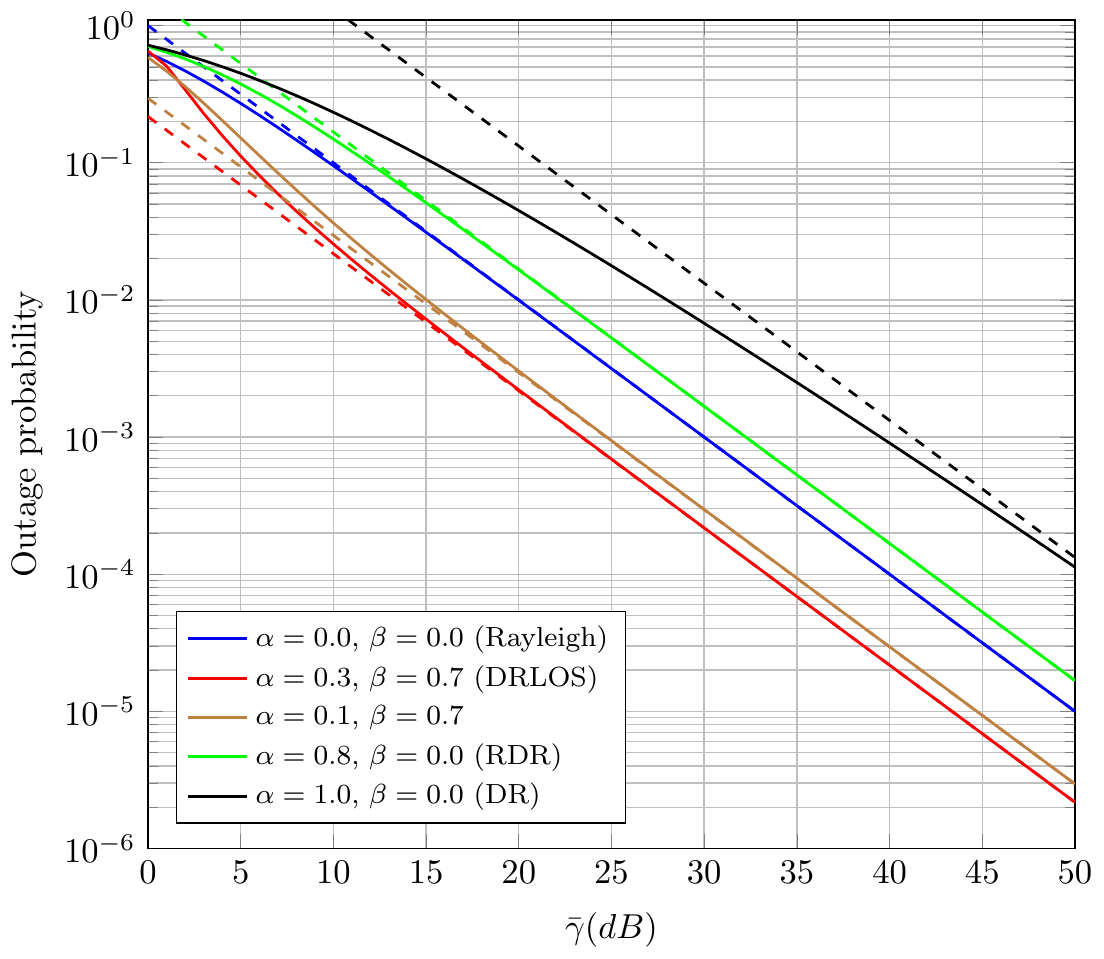}
\centering
\caption{Exact (solid line) and asymptotic (dashed line) outage probability for $\gamma_{th}=0$ dB and different values of $\alpha$ and $\beta$.} 
\label{Fig_Outage}
\end{figure}

This effect is also observed when studying the outage probability, defined as the probability that the instantaneous SNR falls below a given threshold $\gamma_{th}$, which can be directly computed from the CDF as $OP=\Pr\{\gamma<\gamma_{th}\}$. This is illustrated in Fig. \ref{Fig_Outage}, where the OP for for $\beta=0.7$ grows as $\alpha$ is decreased. We also note the tightness of the tail approximations for the CDF given by \eqref{eqnew01}.

\section{Conclusion}
We provided an alternative formulation for the statistics of SOSF channels, which has numerous practical advantages over the original one. By identifying the SOSF channel as a conditional Rician channel averaged over an exponential distribution, simpler expressions for the PDF and CDF of this relevant fading distribution are obtained, together with a novel expression for the MGF. A general way to conduct the performance analysis of wireless communication systems operating over SOSF channels is also introduced. Strikingly, in the presence of a reasonably high LOS component, a double-Rayleigh diffused component is less detrimental for system performance than a Rayleigh-diffused one.
\label{Conclusions}


\appendix
\label{app1}
In this appendix the relation  $\int_{0}^{b}\Gamma(0,x,z)dz=e^{-x}-\Gamma(1,x,b)$ is demonstrated.

\begin{IEEEproof}
 Starting with the definition of the generalized incomplete gamma function 
\begin{equation}
\label{Eq:GGF_def}
\Gamma(a,x,b)=\int_{x}^{\infty}t^{a-1} e^{-t}e^{\frac{-b}{t}}dt,
\end{equation} 
we can write
\begin{equation}
\int_{0}^{b}\Gamma(0,x,z)dz=\int_{0}^{b}\left(\int_{x}^{\infty}t^{-1} e^{-t}e^{\frac{-z}{t}}dt\right) dz.
\end{equation}

Interchanging the order of integration we have

\begin{equation}
\label{Eq:Doble_int_gamma}
\int_{0}^{b}\Gamma(0,x,z)dz=\int_{x}^{\infty}t^{-1} e^{-t} \left(\int_{0}^{b}e^{\frac{-z}{t}} dz\right)dt.
\end{equation}
The inner integral takes the value $t(1-e^{-\frac{b}{t}})$. Substituting in (\ref{Eq:Doble_int_gamma}) we get
\begin{equation}
\int_{0}^{b}\Gamma(0,x,z)dz=\int_{x}^{\infty} e^{-t}dt-\int_{x}^{\infty} e^{-t}e^{\frac{-b}{t}}dt,
\end{equation}
where the second term corresponds to the definition of the generalized incomplete gamma function (\ref{Eq:GGF_def}) with parameter $a=1$, resulting in
\begin{equation}
\int_{0}^{b}\Gamma(0,x,z)dz=e^{-x}-\Gamma(1,x,b).
\end{equation}
\end{IEEEproof}

\bibliographystyle{ieeetr}
\bibliography{SOSF}

\end{document}